\documentclass{article}

\usepackage{arxiv}

\usepackage[utf8]{inputenc} 
\usepackage[T1]{fontenc}    
\usepackage{hyperref}       
\usepackage{url}            
\usepackage{booktabs}       
\usepackage{amsfonts}       
\usepackage{nicefrac}       
\usepackage{microtype}      
\usepackage{lipsum}		
\usepackage{graphicx}
\usepackage{natbib}
\usepackage{doi}

\usepackage{multirow}
\usepackage{hhline}
\usepackage{todonotes}

\usepackage[inline]{enumitem}
\newcommand{\devops}{\textit{DevOps}} 
\newcommand{\etal}{\textit{et al.}}
\newcommand{\ie}{\textit{i.e.}, }
\newcommand{\itemheader}[1]{\textbf{#1}}
\newcommand{\interviewee}[1]{\texttt{#1}}

 \title{\devops{} Adoption: Eight Emergent Perspectives}

\author{ \href{https://orcid.org/0000-0003-2802-6440}{\includegraphics[scale=0.06]{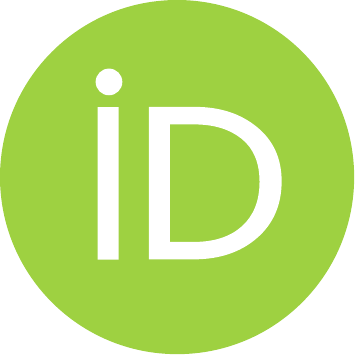}\hspace{1mm}Mauro Lourenço Pedra}\\
	Graduate Program in Informatics\\
	Federal University of Rio de Janeiro (UFRJ)\\
	Rio de Janeiro, Brazil \\
	\texttt{mauro.pedra@ufrj.br} \\
	\And
	\href{https://orcid.org/0000-0003-0951-6612}{\includegraphics[scale=0.06]{orcid.pdf}\hspace{1mm}Mônica Ferreira da Silva} \\
	Graduate Program in Informatics\\
	Federal University of Rio de Janeiro (UFRJ)\\
	 Rio de Janeiro, Brazil \\
	\texttt{monica.silva@ppgi.ufrj.br} \\
	\And
	\href{https://orcid.org/0000-0002-2109-1285}{\includegraphics[scale=0.06]{orcid.pdf}\hspace{1mm}Leonardo Guerreiro Azevedo} \\
	IBM Research\\
    Rio de Janeiro, Brazil \\
	\texttt{lga@br.ibm.com} \\
}

\date{}

\renewcommand{\shorttitle}{\devops{} adoption}

\begin{document}
\maketitle

\begin{abstract}

\devops{} is an approach based on lean and agile principles in which business, development, operations, and quality teams cooperate to deliver software continuously aiming at reducing time to market, and receiving constant feedback from customers. 
However, implementing \devops{} can be a complex and challenging mission due it requires significant paradigm shift. Consequently, many failures and misconceptions can occur about \devops{} adoption by organizations, despite its numerous benefits.

This work identifies, describes, and compares different perspectives related to \devops{} adoption in academy and industry. 
The perspectives can be understood as factors or variables that influence or help to understand the \devops{} journey. 

We employed a sequential multi-method research approach, including Systematic Literature Review (SLR) and Case Study.
As a result, eight perspectives were found: concepts, models, principles, practices, difficulties, challenges, benefits, and strategies. 
More specifically, the SLR produced 390 items, which can be understood as occurrences of a perspective.
The conducted case study confirmed 75 items, corroborating the SLR findings, while another 29 items emerged.
This global view on \devops{} adoption may guide beginners, both theorists, and practitioners, to make the necessary organizational transformation less painful.

\end{abstract}


\section{Introduction}

Several organizations that develop and use information systems split their software teams structurally into departments.
A standard approach is a separation in software development and system operation departments.
This division has been debated, and combining those teams is argued in the concept of \devops{}. \devops{} is about fast, flexible development and provisioning business processes. It efficiently integrates development, delivery, and operations, thus facilitating a lean, fluid connection of these traditionally separated silos~\cite{ebert2016devops}.

Usually, the Development team is responsible for answering market change needs and delivering new assets in production as quickly as possible.
On the other hand, the Operation team is responsible for providing stable, trustful, and secure IT services to clients, ensuring that changes that may compromise the services are not deployed in production~\cite{kim2016devops}.
Therefore, Development and Operation teams have conflicting stimuli and goals due to the choice between agility and stability. So, the industry has adopted \devops{}  in an attempt to deal with this disparate of concerns~\cite{fitzgerald2017continuous}.

Although \devops{} has been an essential movement in industry for more than a decade, it has not received much attention from the academic community until recently~\cite{wiedemann2019devops}. Still, the interest is increasing from both practitioners and researchers.
Among the main reasons are results such as increased productivity and quality of IT products and services and reduced development cycle time, and, consequently, faster response to market needs~\cite{dingsoyr2016emerging}. 
Although many organizations have achieved great success with technology transformations, there is still much work to be done — both in the broader industry and within individual organizations~\cite{humble2018accelerate}.

\devops{} adoption is not an easy or direct task. It may require requires complex changes in an organization, corporate process, and workflows~\cite{bucena2017simplifying}.
As a software process improvement initiative, the route for a successful \devops{} adoption is exclusive and typical of each organization.
However, it is possible to learn from empirical knowledge to plan future \devops{} adoption initiatives~\cite{smeds2015devops}. 
Research on \devops{} is fragmented, making it challenging to identify and understand its scope, covered topics and challenges already addressed as well as those not yet addressed~\cite{erich2014cooperation}.

Organizational and cultural changes and adoption of technologies (mainly frameworks) are required to overcome the detachment of Development and Operation teams~\cite{wettinger2014devopslang}. 
In addition, given that \devops{} practitioner literature is still emerging, and reliable academic research on the phenomenon is sparse, IT organizations lack concrete guidance on how to approach the \devops{} paradigm in practice ~\cite{nielsen2017closing} . 

The goal of this work is to identify, describe, and compare different perspectives related to \devops{} adoption in academy and industry. 
These perspectives refer to possible dimensions or ways to interpret or visualize the phenomenon of \devops{} adoption.
In other words, the perspectives can be understood as factors or variables that influence or help to understand the \devops{} journey. 
We provide a holistic view of these perspectives without delving into any specific characteristics. 

\begin{figure}[h]
	\centering
	\includegraphics[width=0.5\linewidth]{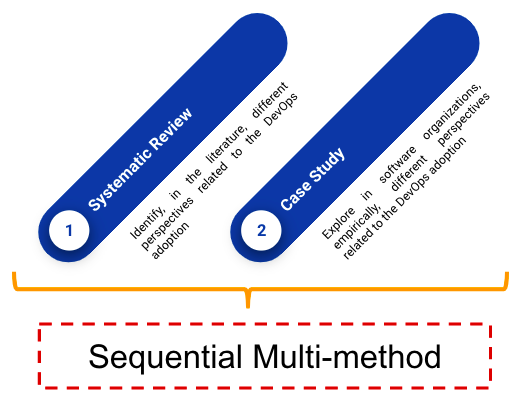}
	\caption{Holistic Research Scheme}
	\label{research-scheme}
\end{figure}

This work employed a sequential multi-method research approach, including Systematic Literature Review (SLR) and Case Study.
Figure~\ref{research-scheme} shows a holistic scheme of the applied multi-method research.
The SLR aimed to identify, describe and deepen the knowledge about \devops{} adoption in the academy and identify the main related perspectives. 
The case study was conducted with two software development organizations and empirically explored how is going \devops{} adoption in industry and compared with the SLR results.
The research was exploratory, descriptive, and deductive since it seeks new insights and aims at generating and validating ideas and hypotheses for future work.
Deductively, according to ~\cite{runeson2012case}, this can lead to confirmed or rejected theories.
More specifically, the SLR produced 390 items, which can be understood as occurrences of a perspective, found in the 28 selected primary studies and classified according to our interpretative scheme. 
The conducted case study confirmed 75 items, corroborating the SLR findings, while another 29 items emerged, expanding what was found in the literature regarding \devops{} adoption. 
This global view on \devops{} adoption may guide beginners, both theorists, and practitioners, to make the necessary organizational transformation less painful.

\section{Systematic Literature Review}

To achieve research objectives, SRL was used as one of the research strategies. This section presents a brief description of the adopted protocol, including selection criteria and discovered limitations, in addition to achieved results.

\subsection{Preliminary Bibliographic Research}

Initially, it was necessary to conduct preliminary bibliographic research of the literature through snowballing method~\cite{wohlin2014guidelines}, \ie{} follow references of the most relevant works in the \devops{} theme to deepen the understanding and construction of a more robust search string.

During this phase, it became evident the difficulty of defining the most common English-speaking terms related to \devops{}. For ensuring more accuracy in the search string's construction, we performed a generic search for the word \devops{} in the search engines \textit{ACM Guide to Computing Literature}, IEEExplore, DBLP, and Scopus. Three hundred forty-one papers were obtained, excluding duplicate results, unavailable publications, and papers written in languages other than English. After that, a file containing all the titles, abstracts, and keywords of the selected publications was generated and submitted to the Corpus linguistic processor\footnote{Available at \url{http://linguistica.insite.com.br/corpus.php}.}, from Insite Linguistics Group, to identify the words with the highest occurrence in the publications. The generated report counted more than 7,000 words. Three stages were used to deal with these words:

\begin{itemize}
	\item Exclusion of irrelevant or generic words such as pronouns, conjunctions, adverbs, and connectives;
	\item Combination of words with similar radicals;
	\item Elimination of words with less than 100 occurrences.
\end{itemize}

\begin{figure}[h]
	\centering
	\includegraphics[width=0.5\linewidth]{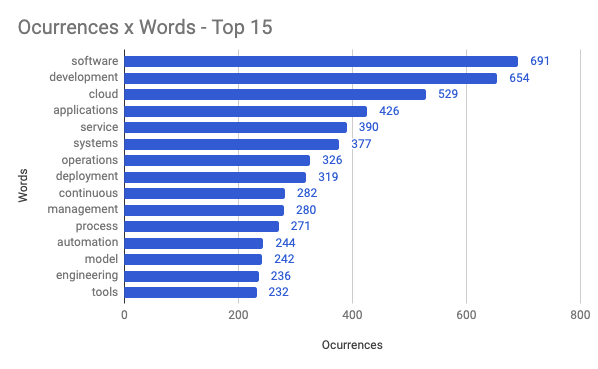}
	\caption{Top 15 of the most relevant words found in \devops{} publications}
	\label{occurrences-words}
\end{figure}

As an example, Figure~\ref{occurrences-words} displays the 15 more relevant words found in the selected publications. We considered the more fitting words to the research objectives to construct the search string, which not necessarily are the more cited ones.

\subsection{Execution}

The following search string was assembled from the preliminary bibliographic research:

\begin{center}
	\textbf{( \devops{} )  AND  ( model  OR  framework  OR  method  OR  platform  OR  solution ) AND ( adoption OR  implementation  OR  introduction  )}
\end{center}

Table~\ref{tab:citacoes-palavras} relates the number of quotes from the words that make up the search string.

\begin{table}
	\caption{Number of Word Citations for Search String}
	\label{tab:citacoes-palavras}
	\begin{minipage}{\columnwidth}
		\begin{center}
			\begin{tabular}{|c|c|}
				\hline
				\textbf{Word} & \textbf{Number of Citations}\\
				\hline
				\devops{} & 1344\\
				\hline
				model & 242\\
				\hline
				framework & 101\\
				\hline
				method & 109\\
				\hline
				platform & 100\\
				\hline 
				solution & 100\\
				\hline
				adoption & 143\\
				\hline
				implementation & 163\\
				\hline
				introduction & 77\\
				\hline
			\end{tabular}
		\end{center}
	\end{minipage}
\end{table}

We chose the following scientific bases among the main ones in Computer Science to combine comprehensive search engines with other more specific bases. This combination aims to achieve greater plurality and coverage of scientific events and journals in the investigated area.

\begin{itemize}
    \item \textbf{Scopus} - \url{www.scopus.com} - Search Engine
    \item \textbf{IEEE Xplore} - \url{ieeexplore.ieee.com} - Bibliographic basis
    \item \textbf{Science Direct} - \url{www.sciencedirect.com} - Bibliographic basis
    \item \textbf{Springer} - \url{link.springer.com} - Bibliographic basis
    \item \textbf{ACM Digital Library} - \url{dl.acm.org} - Hybrid
\end{itemize}

The Mendeley\footnote{Available at \url{https://www.mendeley.com/}, and licensed by UFRJ, the institution of some of the authors.} software was used as a support tool to manage the documents extracted from the scientific bases, generate bibliographies, and add markings in the texts.

\begin{figure}[h]
	\centering
	\includegraphics[width=0.7\linewidth]{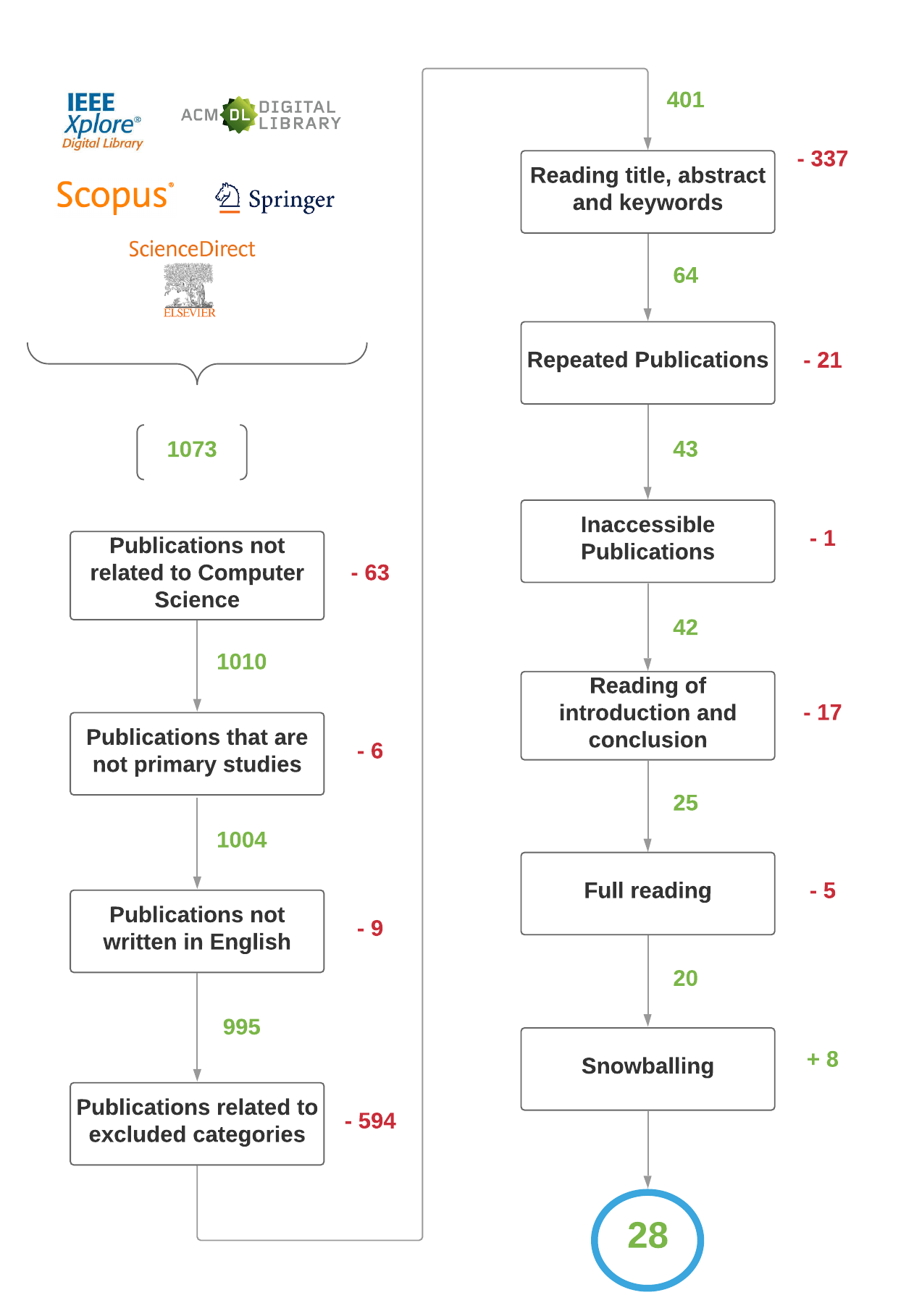}
	\caption{Application of Inclusion and Exclusion Criteria}
	\label{fig:criterio-selecao}
\end{figure}

In this stage of the SLR, the inclusion and exclusion criteria were applied to increase selected studies' relevance, as follows in Figure~\ref{fig:criterio-selecao}.

The Snowballing search was applied to the twenty primary studies initially selected, which added eight new studies, resulting in twenty-eight documents in total. This step was performed to avoid missing relevant works.

\subsection{Analysis and Results}

This section will detail the extraction and organization of information, using spreadsheets, thematic analysis, and mind mapping to present the results. Eight perspectives were generated, \ie{}, concepts, models, principles, practices, difficulties, challenges, benefits, and strategies.

\subsubsection{Concepts}
\label{rsl-conceitos}

To obtain the author's view on \devops{} concepts, we extracted a term definition from each of the 28 studies.

\begin{figure}[h]
	\centering
	\includegraphics[width=0.5\linewidth]{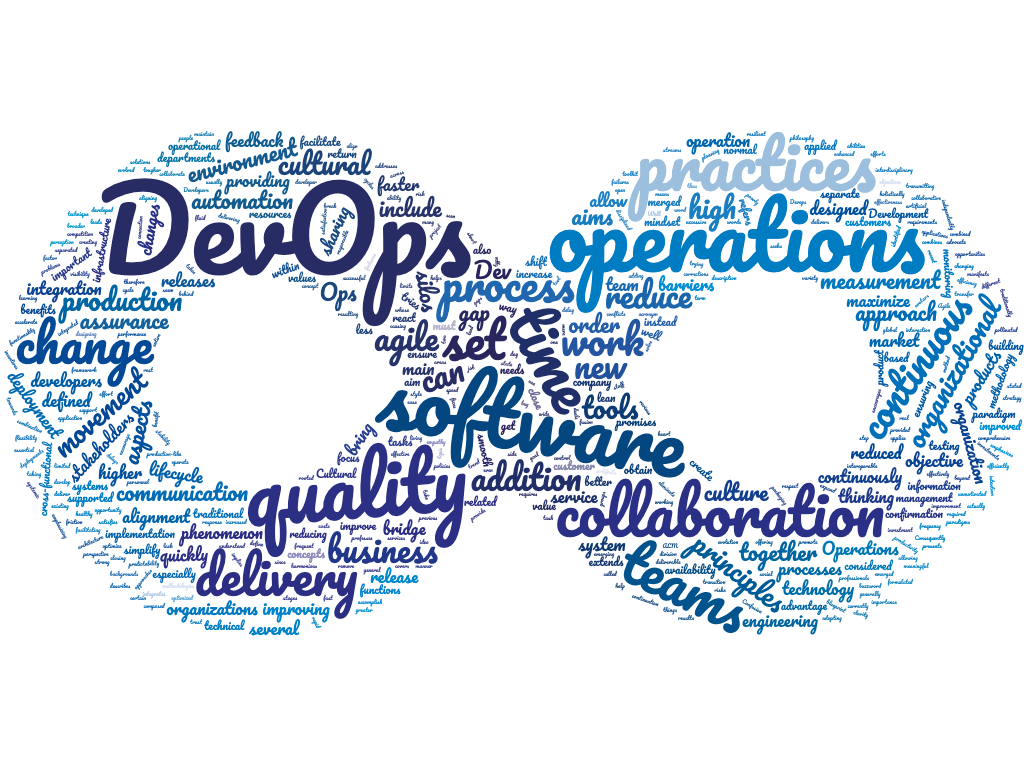}
	\caption{Cloud of Words}
	\label{fig:concept-devops}
\end{figure}

A word cloud was created to provide a more comprehensive view of the various concepts extracted, as shown in Figure~\ref{fig:concept-devops}. The greater the number of words present, the larger the size or highlight shown. Stop words such as pronouns, conjunctions, adverbs, and connectives were not considered.
The site WordClouds.com\footnote{Available at \url{https://www.wordclouds.com/}.} was used to create the word cloud. 

This word cloud reinforces some terms related to \devops{}, such as collaboration, culture, software, practices, quality, delivery, automation, operations, processes, and teams, which shows excellent adherence and convergence to the broad concept \devops{} represents. Moreover, although there are different interpretations in \devops{} literature, this word cloud helps to visualize a connection and many similarities between the various authors considered in this study.

\subsubsection {Distribution of Primary Studies by Locality}
\begin{figure}[h]
	\centering
	\includegraphics[width=0.5\linewidth]{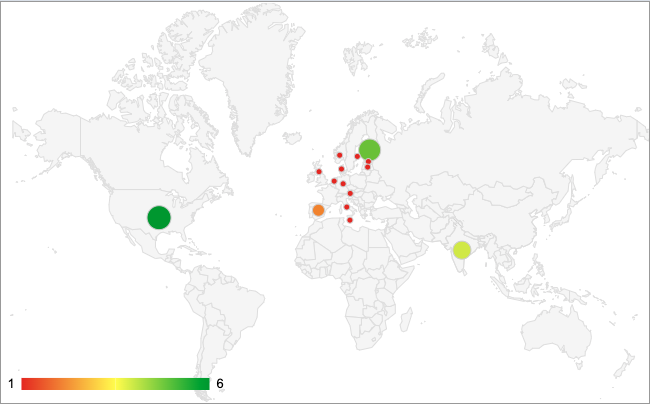}
	\caption{Distribution of Primary Studies Around the World}
	\label{fig:dist-pesquisa}
\end{figure}

As an additional result from the SLR, in Figure~\ref{fig:dist-pesquisa}, it is possible to visualize, through the illustrated world map, the distribution of the primary studies selected in this phase of the research by country of publication.
There is a high concentration in the European and North American continent countries and relevant research in India.
From this, it is possible to deduce that there is a scarcity of qualified studies, specialized events, or research focus in the least developed countries, probably due to a low degree of maturity or interest in \devops{}' adoption. Future research on this topic may be helpful.

\subsubsection{\devops{}-related Models and Frameworks}

Table~\ref{tab:modelos-consolidado} presents a consolidated view of the models and frameworks found in the 28 studies selected in the SLR, as well as their type, focus, and quantity.

\begin{table}
	\caption{\devops{}-related Models and Frameworks - Consolidated}
	\label{tab:modelos-consolidado}
	\begin{minipage}{\columnwidth}
		\begin{center}
			\begin{tabular}{|c|c|c|}
				\hline
				\textbf{Type} & \textbf{Focus} & \textbf{Quantity}\\
				\hline
				\multirow{3}{*}{Model} & \devops{} Adoption & 4 \\
				\hhline{~--} & \devops{} Maturity & 2 \\
				\hhline{~--} & Technology Acceptance & 1 \\
				\hline
				\multirow{4}{*}{Framework} & \devops{} Adoption & 2 \\
				\hhline{~--} & \devops{} Maturity & 2 \\
				\hhline{~--} & Knowledge Sharing & 1 \\
				\hhline{~--} & Communication & 1 \\
				\hline
			\end{tabular}
		\end{center}
	\end{minipage}
\end{table}

\subsubsection{Thematic Analysis and Mind Mapping}

Although the kind of results obtained by content analysis techniques cannot be taken as irrefutable proof, it constitutes an illustration that allows corroborating, at least partially, the presuppositions in question~\cite{bardln1977analise}.

Carrying out a thematic analysis consists of discovering the ``nuclei of meaning'' that make up the communication and whose presence, or frequency of appearance, can mean something for the chosen analytical objective. The theme is generally used as a recording unit to study motivations for opinions, attitudes, values, beliefs, and trends. In this sense, it is possible to consider that the mind mapping technique can be seen as a type of content analysis, even more, a kind of thematic analysis.

Mind maps were chosen because they can represent ideas linked to a central theme. There are a lot of rules for creating mind maps in which the main one is to bring the brain in and use the imagination~\cite{crowe2012mind}. Creating mind maps is an easy and natural organization method and visualizing complex data, both as survey methods and as interactions between data. In addition, mind maps can also help people learn concepts better than linear formats and annotations.

Given the large number of occurrences associated with each perspective, it was necessary to adopt a classification scheme using graduated weights. This scheme considers an importance degree (or weight) for each item present in each perspective, from the association of a star that can have three colors (or three weights): \textit{red} represents the highest degree, with value three; \textit{yellow} has value two; and, finally, \textit{blue} represents the lowest degree, with value one. This degree is determined by the number of times the item is cited, considering one citation if the item appears one or more times in the same primary study. On each picture that represents a perspective's mind map, there is a legend that explains this particular color scheme. The number of citations present in this color scheme may vary according to each perspective, never exceeding the limit of 28 citations, which is the total number of primary studies considered.

In addition, given the possibilities and complexities inherent in the mind map's construction, it was necessary to create a type called aggregator branch (expressed by a noun). This aggregator refers to a branch that explains, details, or exemplifies the branch or avoids repetition of similar terms, improving semantics. An image of a green check sign represents the aggregator, and a legend is associated with it.

An additional feature of the classification scheme was still necessary to better express the mind maps' semantics in scenarios of greater complexity and number of items. This scheme was applied in the mind maps of the \textit{Difficulties, Challenges, Benefits, and Strategies} perspectives. To represent the degree of importance of a mind map branch, the sum of the weights attributed to all the branch twigs and leaves, and sub-leaves was considered. This number was represented below the box that represents the topic itself. The number that appears next to the topic is the sum of the number of subordinate items. It is important to remember that the values assigned to the weights always vary from 1 to 3, with an increasing degree of importance. The criterion for framing an item to a given weight is specified in the legend of each mind map. It may vary from perspective to perspective, according to the free criterion adopted by the authors.

\begin{figure}[h]
	\centering
	\includegraphics[width=0.7\linewidth]{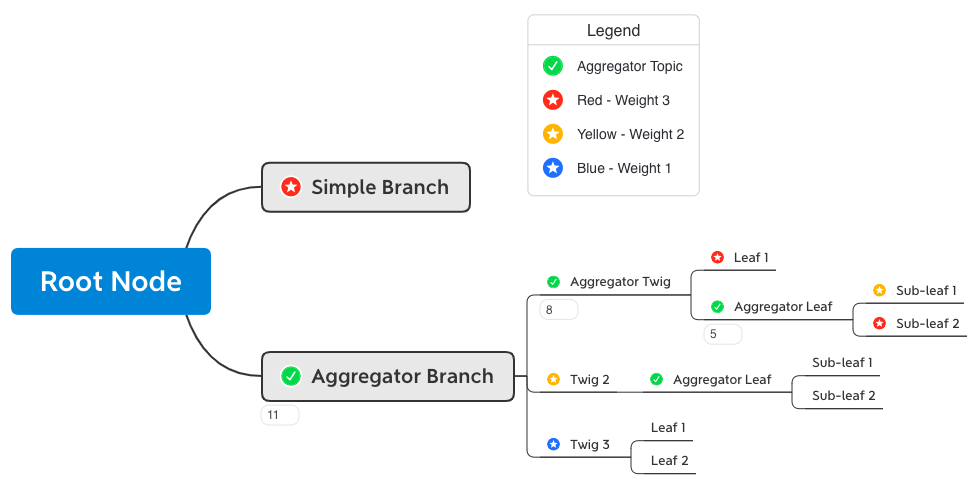}
	\caption{Structural Meta-Model of Mind Maps}
	\label{fig:meta-model}
\end{figure}

To facilitate the visualization and understanding of this internal classification scheme of mind maps, a structural meta-model was created, represented in  Figure~\ref{fig:meta-model}, in which:

\begin{itemize}
	\item Root Node = Perspective
	\item Branch = Correlated Item or Aspect
	\item Twigs and Leaves = Items or Aspects' aggregations or extensions
\end{itemize}

Given the possible options in drawing a mind map, some clarifications about construction rules are in order. A simple branch directly represents an item or aspect related to the perspective. At most, it can have a few associated branches, only at 1 level of depth, to explain it or correlate similar items or aspects. 

On the other hand, the aggregator branch or aggregator topic, present in several perspectives, represents a set of correlated items or aspects. It can contain another branch, a twig, leaf, or sub-leaf. Its existence allows for greater freedom in drawing the mind map, unifying or simplifying several items correlated by some characteristic to improve the semantics and contain any reasonable explanation found in primary studies. It can be considered a complex branch that represents high-density structures with several branches. The main item or aspect, which represents something related to the perspective and which must be considered in the accounting, in any branch, will always have an associated colored star.

The XMind ZEN\footnote{\url{https://www.xmind.net/}} software tool was used to support the creation of mind maps. 
Due to the extensive size of mind maps, we provided a Web page\footnote{\url{https://github.com/leogazevedo/devops-adoption-perspectives}} with them.

\subsubsection{Principles}

We found fifteen principles related to \devops{}. Among the five most cited principles are only two of those present in the acronym CALMS, they are: Culture and Automation. Among the most cited are the principles of Communication, Collaboration and Agility. The other principles present in the acronym CALMS (Lean, Measurement and Sharing) also appear, but with fewer citations in the selected studies, thus configuring themselves with less relevance in this research.

In Table~\ref{tab:desc-principles}, there is a description of each principle found in the 28 selected studies, as well as its relevance. This relevance was calculated considering the number of citations, as described below:

\begin{itemize}
	\item High = Over 19 occurrences
	\item Medium = Among 10 and 19 occurrences
	\item Low = Under 10 occurrences
\end{itemize}

\begin{table}[]
\centering
\caption{Principles}
\label{tab:desc-principles}
\begin{tabular}{|c|c|c|}
	\hline
	ID &
	Principle &
	Relevance \\ \hline
	\begin{tabular}[c]{@{}c@{}}P1\end{tabular} &
	\begin{tabular}[c]{@{}c@{}}Collaboration\end{tabular} &
		\begin{tabular}[c]{@{}c@{}}High\end{tabular}  \\ \hline
	\begin{tabular}[c]{@{}c@{}}P2\end{tabular} &
	\begin{tabular}[c]{@{}c@{}}Automation\end{tabular}  & 
	\begin{tabular}[c]{@{}c@{}}High\end{tabular}  
	  \\ \hline
	\begin{tabular}[c]{@{}c@{}}P3\end{tabular} &
	\begin{tabular}[c]{@{}c@{}}Culture\end{tabular} &
	\begin{tabular}[c]{@{}c@{}}High\end{tabular} 
	\\ \hline
	\begin{tabular}[c]{@{}c@{}}P4\end{tabular} &
	Communication &
	\begin{tabular}[c]{@{}c@{}}High\end{tabular} \\ \hline
	\begin{tabular}[c]{@{}c@{}}P5\end{tabular} &
	\begin{tabular}[c]{@{}c@{}}Agility\end{tabular} &
	\begin{tabular}[c]{@{}c@{}}High\end{tabular} \\ \hline
	\begin{tabular}[c]{@{}c@{}}P6\end{tabular} &
	Sharing &
	\begin{tabular}[c]{@{}c@{}}Medium\end{tabular} \\ \hline
	\begin{tabular}[c]{@{}c@{}}P7\end{tabular} &
	Measurement &
	Medium \\ \hline
	\begin{tabular}[c]{@{}c@{}}P8\end{tabular} &
	\begin{tabular}[c]{@{}c@{}}Trust\end{tabular} &
	\begin{tabular}[c]{@{}c@{}}Medium\end{tabular} \\ \hline
	\begin{tabular}[c]{@{}c@{}}P9\end{tabular} &
	\begin{tabular}[c]{@{}c@{}}Lean\end{tabular} &
	\begin{tabular}[c]{@{}c@{}}Medium\end{tabular} \\ \hline
	\begin{tabular}[c]{@{}c@{}}P10\end{tabular} &
	\begin{tabular}[c]{@{}c@{}}Alignment of \\ Responsabilities\end{tabular} &
	\begin{tabular}[c]{@{}c@{}}Medium\end{tabular} \\ \hline
	\begin{tabular}[c]{@{}c@{}}P11\end{tabular} &
	\begin{tabular}[c]{@{}c@{}}Respect\end{tabular} &
	\begin{tabular}[c]{@{}c@{}}Low\end{tabular} \\ \hline
	\begin{tabular}[c]{@{}c@{}}P12\end{tabular} &
	\begin{tabular}[c]{@{}c@{}}Transparency\end{tabular} &
	\begin{tabular}[c]{@{}c@{}}Low\end{tabular} \\ \hline
	\begin{tabular}[c]{@{}c@{}}P13\end{tabular} &
	\begin{tabular}[c]{@{}c@{}}Resilience\end{tabular} &
	\begin{tabular}[c]{@{}c@{}}Low\end{tabular} \\ \hline
	\begin{tabular}[c]{@{}c@{}}P14\end{tabular} &
	\begin{tabular}[c]{@{}c@{}}Standardization\end{tabular} &
	\begin{tabular}[c]{@{}c@{}}Low\end{tabular} \\ \hline
	\begin{tabular}[c]{@{}c@{}}P15\end{tabular} &
	\begin{tabular}[c]{@{}c@{}}Collective Property\end{tabular} &
	\begin{tabular}[c]{@{}c@{}}Low\end{tabular} \\ \hline
\end{tabular}
\end{table}

\subsubsection{Practices}

\begin{figure}[h]
	\centering
	\includegraphics[width=0.5\linewidth]{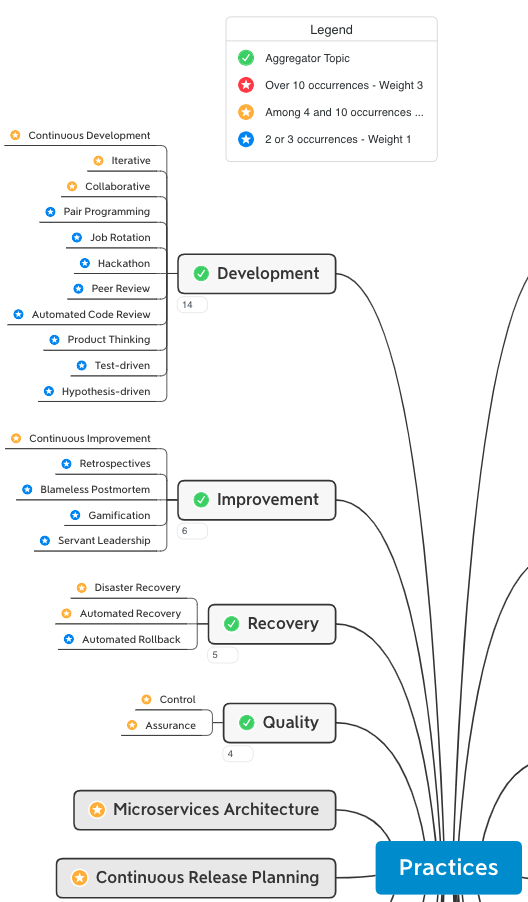}
	\caption{Mind Map of \devops{} Practices - Part 1}
	\label{fig:practices-1}
\end{figure}

\begin{figure}[h]
	\centering
	\includegraphics[width=0.5\linewidth]{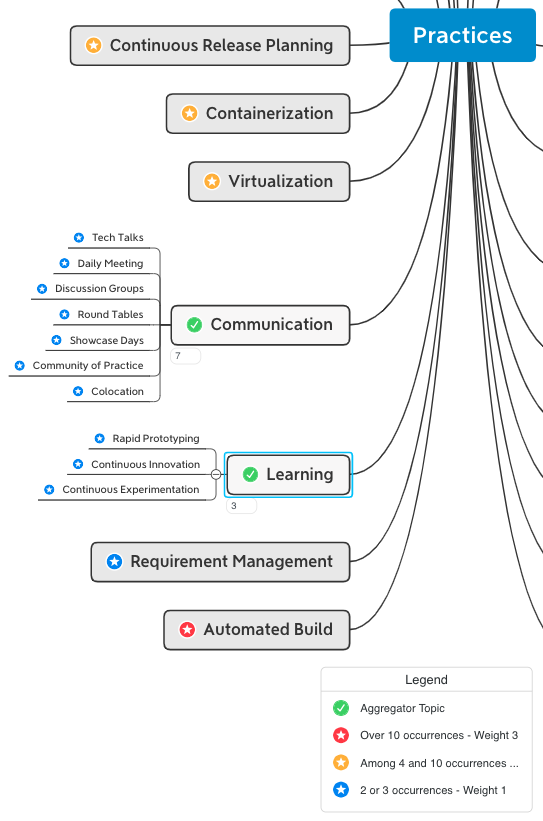}
	\caption{Mind Map of \devops{} Practices - Part 2}
	\label{fig:practices-2}
\end{figure}

\begin{figure}[h]
	\centering
	\includegraphics[width=0.7\linewidth]{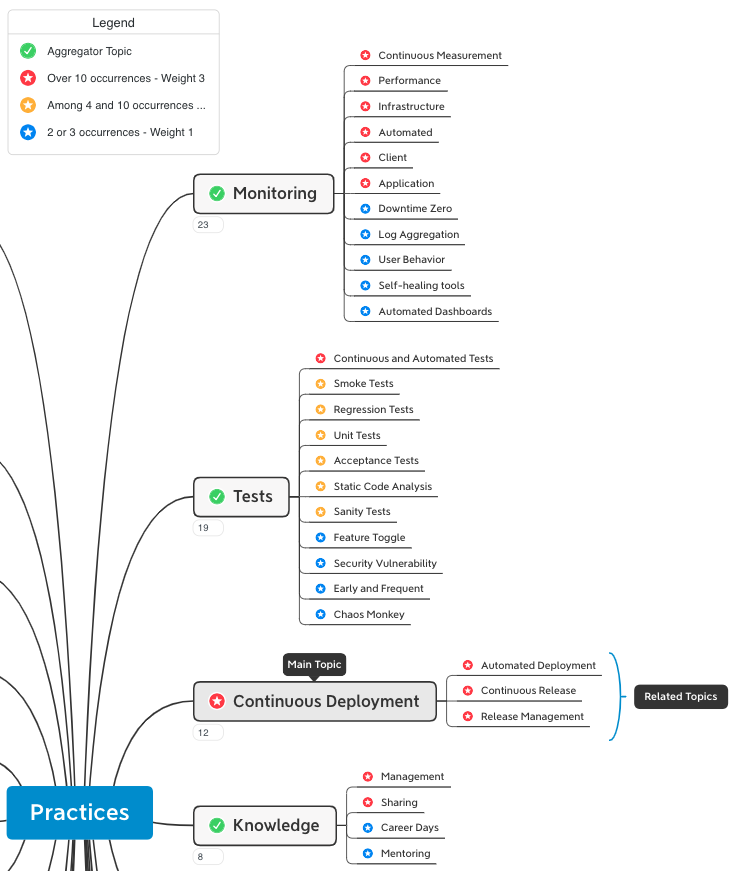}
	\caption{Mind Map of \devops{} Practices - Part 3}
	\label{fig:practices-3}
\end{figure}

\begin{figure}[h]
	\centering
	\includegraphics[width=0.7\linewidth]{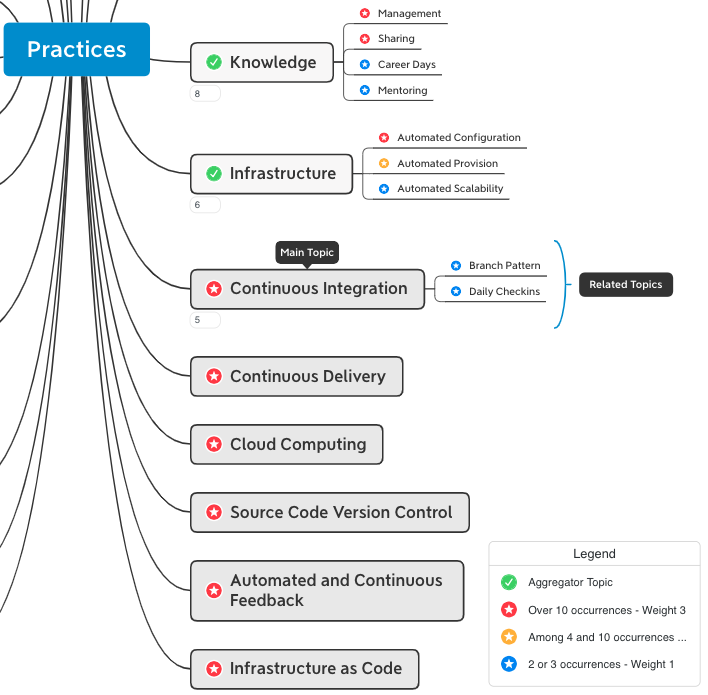}
	\caption{Mind Map of \devops{} Practices - Part 4}
	\label{fig:practices-4}
\end{figure}

Eleven direct practices were found represented by simple branches, two indirect practices represented by hybrid branches, and ten aggregator branches were created to facilitate the organization and visualization, as they were considered to represent correlated practices. Aggregator branches represent the number of practices that are grouped, making a subtotal of 60 practices. That is, in total, adding the direct practices (11), plus the indirect practices (2) with the practices present in the aggregator branches (60), 73 practices related to \devops{} were found. Practices related to Monitoring, Development, Testing, and Continuous Deployment stand out among the most cited. As a demonstration, the \devops{} Practices mind map was represented in 4 parts, for ease of viewing, due to its length, as we can see in the figures~\ref{fig:practices-1},~\ref{fig:practices-2},~\ref{fig:practices-3} and~\ref{fig:practices-4}. 
The most cited practices with great relevance (red star) to the \devops{} concept are:

\begin{itemize}
    \item Automated Build
    \item Continuous Deployment
    \item Continuous Integration
    \item Continuous Delivery
    \item Cloud Computing
    \item Source Code Version Control
    \item Automated and Continuous Feedback
    \item Infrastructure as Code
\end{itemize}

\subsubsection{Difficulties}

Forty difficulties were encountered, represented by five aggregator branches. These were created to better semantically represent difficulties related to Tools (4), Client (2), Teams (14), Management (8), and Process (12). The weights of the classification scheme highlight difficulties or problems related to the Team and Process aspects during the adoption of \devops{}.

\subsubsection{Challenges}

Fifty-two challenges were found, represented by four aggregator branches. These were created to better semantically represent challenges related to Tools (4), Teams (11), Management (22), and Process (15). Applying the weights of the classification scheme, challenges or risks related to the Management and Process aspects during the adoption of \devops{} are highlighted.

\subsubsection{Benefits}

Seventy benefits were found, represented by six aggregator branches. These were created to better semantically represent benefits related to Product/Service (9), Customer (8), Teams (19), Market (1), Organization (10), and Process (23). The weights of the classification scheme highlight benefits related to the aspects of the Process and Teams during the adoption of \devops{}.

\subsubsection{Strategies}

Ninety-nine strategies were found, represented by four aggregator branches. These were created to better semantically represent strategies related to Organization (14), Teams (27), Management (18), and Process (40). The weights of the classification scheme highlight strategies related to the aspects of the Process and Teams during the adoption of \devops{}.

\subsubsection{Summary of Perspectives}

To facilitate the comparison between perspectives and consolidate the results achieved, the tables~\ref{tab:perspec-sem-peso} and~\ref{tab:perspec-com-peso} were created. These tables present four perspectives and the analyzed branches or dimensions. The difference between the tables occurs regarding the use or not of the weights of the classification scheme. The use of weights did not change the order or degree of importance either of the perspectives, comparative scope, or branches or dimensions when viewed in isolation and compared to each other. So, using weights helps reinforce the validity and assertiveness of using the mind mapping technique to encode information extracted from selected studies.

For the \textit{Principles} and \textit{Practices} perspectives, there was no need to use specific branches or dimensions due to the ease to analyze and condense the results because of the small number of extracted items, the direct mentions, or good correlation to the perspectives in the analyzed primary studies. For these reasons, these 2 perspectives do not appear in the tables~\ref{tab:perspec-sem-peso} and~\ref{tab:perspec-com-peso}.
			
	\begin{table}[]
		\centering
		\caption{Summary of Perspectives - Weightless}
		\label{tab:perspec-sem-peso}
		\begin{tabular}{|c|c|c|c|c|c|}
			\hline
			& Difficul. & Challen. & Benef.  & Strateg. & Total       \\
			\hline
			Tools     & 4            & 4        & N/A         & N/A         & 8           \\
			\hline
			Manag.          & 8            & 22       & N/A         & 18          & \textbf{48} \\
			\hline
			Teams           & 14           & 11       & 19          & 27          & \textbf{71} \\
			\hline
			Process        & 12           & 15       & 23          & 40          & \textbf{90} \\
			\hline
			Client         & 2            & N/A      & 8           & N/A         & 10          \\
			\hline
			Organiz.     & N/A          & N/A      & 10          & 14          & 24          \\
			\hline
			Product & N/A          & N/A      & 9           & N/A         & 9           \\
			\hline
			Market         & N/A          & N/A      & 1           & N/A         & 1           \\
			\hline
			Total           & 40           & 52       & \textbf{70} & \textbf{99} &  \\
			\hline          
		\end{tabular}
	\end{table}
	
	\begin{table}[]
		\centering
		\caption{Summary of Perspectives - Weighted}
		\label{tab:perspec-com-peso}
		\begin{tabular}{|c|c|c|c|c|c|}
			\hline
			& Difficul. & Challen & Benef.       & Strateg.      & Total        \\
			\hline
			Tools & 6        & 10       & N/A          & N/A          & 16           \\
			\hline
			Manag.      & 16       & 31       & N/A          & 50           & \textbf{97}  \\
			\hline
			Teams       & 24       & 24       & 32           & 63           & \textbf{143} \\
			\hline
			Process    & 18       & 24       & 31           & 83           & \textbf{156} \\
			\hline
			Client     & 2        & N/A      & 11           & N/A          & 13           \\
			\hline
			Organiz. & N/A      & N/A      & 14           & 37           & 51           \\
			\hline
			Product     & N/A      & N/A      & 15           & N/A          & 15           \\
			\hline
			Market     & N/A      & N/A      & 3            & N/A          & 3            \\
			\hline
			Total       & 66       & 89       & \textbf{106} & \textbf{233} &          \\  
			\hline 
		\end{tabular}
	\end{table}

\section{Case Study}

This section presents the case study which was conducted following Runeson \etal{} protocol~\cite{runeson2012case} as follows:

\begin{itemize}
	\item \itemheader{Case study type}: 
	    Exploratory, descriptive, multiple, and holistic.
	
	\item \itemheader{Goal}: 
	    Analyze the state of the practice comparing with state of the art (\ie{} considering SLR results), aiming at bringing new visions, contexts, perspectives, and interpretations enhancing the theory.
	    
	\item \itemheader{The case}: 
	    Professional experts in \devops{} of organizations that handle Software Engineering problems.
	 
	\item \itemheader{Selection strategy}: 
	    Enterprises that are adopting or are mature in \devops{}, even partially, and utilize agile or hybrid methodologies for software development in at least one year.
	 
	 \item \itemheader{Data collection}: through semi-structure, remote interviews, preferably with audio and video recording consented by the interviewees.
	
	\item \itemheader{Quantity of interviewees}: Four experts were selected considering experience, knowledge, and human factors (relationship).
    
    \item \itemheader{Validation (pre-test)}: with a member of HumânITas research group\footnote{Research group of Graduate Program in Informatics of UFRJ (PPGI-UFRJ) which is composed of Ph.D., masters, and students. The group focus is research on human factors about development, adoption, and use of IT. More information available at \url{http://dgp.cnpq.br/dgp/espelhogrupo/635926}.}
    
    \item \itemheader{Interview duration}: About 120 minutes on average.

\end{itemize}

First, a semi-structured interview questionnaire with open and closed questions was created based on SLR results\footnote{Available at \url{https://github.com/leogazevedo/devops-adoption-perspectives}}.
The questionnaire was enhanced considering the feedback of HumânITas research group, and it was composed of three parts:

\begin{itemize}
    \item \itemheader{Social-demographic questions}:
        Questions to characterize the individuals and organization with a focus on the case;
    
    \item \itemheader{Technical questions}: 
        technical questions about \devops{} centered on this research goal.    
    
    \item \itemheader{Practices knowledge form}: a set of practices and resources found in the literature to validate their representativeness and importance in a practical environment.
    
\end{itemize}


Four interviews were conducted by Skype\footnote{\url{https://www.skype.com/}} with four software engineering professions of two distinct organizations.
The same set of questions were applied to all interviewees, and other spontaneous questions were made when necessary. 
The interviewees received disclosure information, and they were asked to present genuine and detailed responses. 
All interview audios were recorded for data collection with the consent of all interviewees.

Table~\ref{tab:entrevistas} presents interview details.
The four interviewees are anonimized with \interviewee{I1}, \interviewee{I2}, \interviewee{I3}, and \interviewee{I4} codes.

\begin{table}[h]
	\caption{Interviews Conduction}
	\label{tab:entrevistas}
	\begin{minipage}{\columnwidth}
		\begin{center}
			\begin{tabular}{|c|c|c|c|}
				\hline
				\textbf{ID} & \textbf{Organization} & \textbf{Date} & \textbf{Duration}\\
				\hline
				\interviewee{I1} & A & May 30th, 2020 & 72 minutes\\
				\hline
				\interviewee{I2} & A & May 30th, 2020 & 125 minutes\\
				\hline
				\interviewee{I3} & B & June 22nd, 2020 & 133 minutes\\
				\hline
				\interviewee{I4} & B & June 22nd, 2020 & 97 minutes\\
				\hline
			\end{tabular}
		\end{center}
	\end{minipage}
\end{table}

Table~\ref{tab:demog-entrev-pes} and Table~\ref{tab:demog-entrev-org} present the demographic data of the interviewees in a personal and organizational context, respectively. 
Table~\ref{tab:desc-empresas} presents the context of the enterprises considering the case study goal.

\begin{table}[h]
\small
	\caption{Interviewee Demographics - Personal Context}
	\label{tab:demog-entrev-pes}
	\begin{minipage}{\columnwidth}
		\begin{center}
			\begin{tabular}{|c|c|c|c|}
				\hline
				\textbf{ID} & \textbf{Gender} & \textbf{Age (years)} & \textbf{Qualification} \\
				\hline
				\interviewee{I1} & Female & 19 to 29 & Incomplete undergrad.\\
				\hline
				\interviewee{I2} & Male & 19 to 29 & Incomplete undergrad.\\
				\hline
				\interviewee{I3} & Male & 30 to 39 & Complete undergrad.\\
				\hline
				\interviewee{I4} & Male & 40 to 49 & Complete graduation\\
				\hline
			\end{tabular}
		\end{center}
	\end{minipage}
\end{table}

\begin{table}[]
\small
	\centering
	\caption{Interviewee Demographics - Organizational Context}
	\label{tab:demog-entrev-org}
	\begin{tabular}{|c|c|c|c|c|}
		\hline
		\textbf{ID} &
		\textbf{Function} &
		\textbf{\begin{tabular}[c]{@{}c@{}}Org.\\ Size\end{tabular}} &
		\textbf{\begin{tabular}[c]{@{}c@{}}IT \\ Experience\end{tabular}} &
		\textbf{\begin{tabular}[c]{@{}c@{}}\devops{} \\ Experience\end{tabular}} \\ \hline
		\interviewee{I1} & \begin{tabular}[c]{@{}c@{}}Product \\ Owner\end{tabular}                                                       & Small  & 2 years                                                  & 1 year   \\ \hline
		\interviewee{I2} & \begin{tabular}[c]{@{}c@{}}Project \\ Director\end{tabular}                                                       & Small  & \begin{tabular}[c]{@{}c@{}}3 years\\ and a half\end{tabular} & 6 months \\ \hline
	 	\interviewee{I3} & \begin{tabular}[c]{@{}c@{}}Business\\ Architect\end{tabular}         & Big   & 12 years                                                 & 9 years  \\ \hline
		\interviewee{I4} & \begin{tabular}[c]{@{}c@{}}Construction \\ Solution \\ Leader\end{tabular}  & Big   & 20 years                                                 & 3 years  \\ \hline
	\end{tabular}
\end{table}

\begin{table}[]
\centering
\caption{Company Description}
\label{tab:desc-empresas}
\begin{tabular}{|c|c|c|}
	\hline
	&
	Company A &
	Company B \\ \hline
	\begin{tabular}[c]{@{}c@{}}Business \\ Branch\end{tabular} &
	\begin{tabular}[c]{@{}c@{}}Provision of \\ IT Services\end{tabular} &
		\begin{tabular}[c]{@{}c@{}}Energy \\ Industry\end{tabular}  \\ \hline
	\begin{tabular}[c]{@{}c@{}}Number of \\ Employees\end{tabular} &
	\begin{tabular}[c]{@{}c@{}}Between \\ 30 and 40\end{tabular}  & 
	\begin{tabular}[c]{@{}c@{}}Between \\ 48 and 50 thousand\end{tabular}  
	  \\ \hline
	\begin{tabular}[c]{@{}c@{}}Locality / \\ Coverage\end{tabular} &
	\begin{tabular}[c]{@{}c@{}}Rio de Janeiro / \\ Regional\end{tabular} &
	\begin{tabular}[c]{@{}c@{}}Brazil / \\ National\end{tabular} 
	\\ \hline
	\begin{tabular}[c]{@{}c@{}}Organizational\\ Structure\end{tabular} &
	Projected &
	\begin{tabular}[c]{@{}c@{}}Functional\\ (matrix efforts)\end{tabular} \\ \hline
	\begin{tabular}[c]{@{}c@{}}IT\\ infrastructure\end{tabular} &
	\begin{tabular}[c]{@{}c@{}}	On-premise\\ (VPS)\end{tabular} &
	\begin{tabular}[c]{@{}c@{}}On-premise \\ (Datacenters)\end{tabular} \\ \hline
	\begin{tabular}[c]{@{}c@{}}Strategic Position\\ IT\end{tabular} &
	Core Business &
	\begin{tabular}[c]{@{}c@{}}Executive \\ Management\end{tabular} \\ \hline
	\begin{tabular}[c]{@{}c@{}}\devops{}\\ Adoption Stage\end{tabular} &
	Initial &
	In progress \\ \hline
\end{tabular}
\end{table}

\subsection{Results and Analysis}

\begin{figure}[h]
	\centering
	\includegraphics[width=0.7\linewidth]{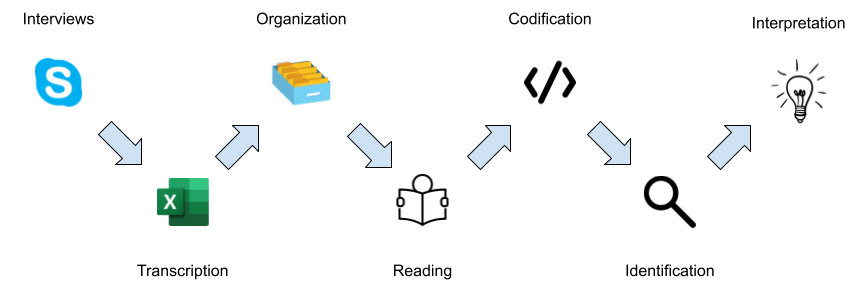}
	\caption{Thematic Analysis Stages}
	\label{fig:thematic-analysis}
\end{figure}

This section presents the results and analysis of the interviews.
We preformed a thematic analysis and compared with SLR results, through the following steps, depicted in Figure~\ref{fig:thematic-analysis}:
\begin{enumerate*}[label=(\roman*)]
    \item Transcription;
    \item Organization;
    \item Reading;
    \item Codification;
    \item Theme identification;
    \item Theme meaning interpretation.
\end{enumerate*}

All interview audio-recorded data were transcribed in a spreadsheet with specific tabs for each interview script stage. 
The interviewee responses were transcribed in sentences connected to the question goal, and redundant and irrelevant terms were removed.
We assigned a unique identification to avoid shuffle and facilitate data analysis.
We exhaustively read the transcriptions to be familiarized and have a clear comprehension of the data.
The codes in each transcription were assigned, analyzing each one manually.
The identified codes were classified into themes according to the perspectives identified in the SLR and their topics.
The frequency and relevance of a code were the main criteria to consider.
In this work, we defined  \textit{frequency} as answers from more than one interviewee and \textit{relevancy} as the importance for the literature and this research.

\subsubsection{Concepts}

For individuals \interviewee{I1} and \interviewee{I2}, from the same company, the focus is on gaining time and productivity.
Therefore, in this company, \devops{} is related to the ideas of agility and \textit{Lean} applied to infrastructure activities.
For individuals \interviewee{I3} and \interviewee{I4}, from the same organization, it is possible to infer a significant focus on collaboration and delivering value to the customer as motivators of the \devops{} initiative.

\subsubsection{Models and Frameworks}

Only \interviewee{I3} reported the existence of a model focused on organizational topology questions tailored for development and operation team\footnote{Available at \url{https://web.devopstopologies.com/}}.

However, in general, interviewees stated that there is no single model or \textit{framework} recognized and widely used as a reference.
It is a journey where each organization must seek its path without following something strictly predetermined since each company has its strategic objectives, maturity, and sizes.

They also mentioned that the multidisciplinary working groups are a good choice for \devops{} adoption.

Besides, \interviewee{I4} suggested the \devops{} adoption through an incremental and on-demand approach using a customized model.
That way, the journey is more adaptable to the different business requirements of each client and technology. 

\subsubsection{Principles}

The two interviewees from company A cited specific \devops{} practices.
Thus, from the items mentioned earlier, it was possible to infer that the adoption of \devops{} in this company had a strong foundation in the automation of infrastructure activities, focusing on implementing and improving the software delivery pipeline.
For these two interviewees, the principles of Automation, in the first place, and Culture, in the second, present themselves as the most important ones related to the adoption of \devops{}.

For the second company, despite the two respondents having different technical expertise, there was a convergence on the most relevant principles considered during the adoption of \devops{}, which can be seen as a positive aspect.
The Culture and Automation principles are also listed as the most relevant in the view of these interviewees.

Therefore, although the two companies and their employees have very different expertise, the principles of Automation and Culture are considered the most relevant in the adoption of \devops{}.
We observe that these principles have a more tangible and visible character since automation brings several changes and practical improvements to the flow and form of work.
In addition, Culture is a principle related to people and has a fundamental role when transforming processes and routines in a way that is as significant as \devops{} proposes.
However, interviewee \interviewee{I3}, depicted that the adoption of \devops{} requires considering all the principles since they are necessary and complementary.
Adopting part of them can bring difficulties and, of course, internal resistance to the changes that come about.

\subsubsection{Practices}

Interviewee cited eleven \devops{} in total.
\interviewee{I3} cited two new practices: InnerSource and Skunkworks.
InnerSource allows software developers to contribute to other teams' efforts, promoting transparency and openness to other people's contributions.
This practice embodies a philosophy of human relations, an approach for rewards and motivations, and a flexible and adaptable set of tools and practices.
Skunkworks project is one developed by a relatively small and loosely structured group of people who research and develop a project primarily for the sake of radical innovation. 
Additionally, \textit{Blue Green Deployment} was mentioned by \interviewee{I1}.

The Continuous Integration practice was the only one cited and recognized by all the interviewees.
Continuous Deployment and Continuous Integration were practices cited by three of the four interviewees.

Regarding the companies, \interviewee{Company A} focuses on automation practices, while \interviewee{Company B} concentrates on automation and is planning to use Cloud Computing.
Seven of the eleven cited practices are related to automation.
This fact reinforces that, in industry, the \devops{} adoption is strongly related to activities of automation practices.



\subsubsection{Difficulties}

\interviewee{I1} and \interviewee{I2} reported difficulties in writing tests, choosing proper tools, adopting this practice broadly in the organization.

\interviewee{I4} highlighted the complexity concerning task automation securely and the human factor, mainly related to convincing and getting team confidence in the new way of working.

\subsubsection{Challenges}

The interviewees of \interviewee{Company A} point out learning and knowledge propagation as a big challenge, which is a fundamental issue for \devops{} adoption. One of its basic principles, as RSL confirmed.

\interviewee{I3} highlighted Cloud Computing as a prerequisite for \devops{} adoption, and the convincing of high management about \devops{} adoption may be pretty costly and comprising.

\interviewee{I4} points out the challenge already taken but not yet overcome about unique open-source tool adoption for code versioning and the internal challenge about resistance to change.

\subsubsection{Benefits}

\interviewee{Company A} has the increase in delivery speed and improvement of company image in the market as the benefits.

\interviewee{I3} reported delivery speed as a benefit resulted from the \devops{} adoption.

Concerning agility increase and workflow simplification, \interviewee{I4} highlighted that process automation reduced the need to take different teams to answer the company requests registered in task control systems.

\subsubsection{Strategies}

\interviewee{I1} and \interviewee{I2} pointed out the requirement to set up a plan for \devops{} adoption, as well as knowledge acquisition and dissemination in an integrated way for the whole organization.
So, there is a concern and focus on stimulating the collaborative culture in the company, which is one of the biggest motivations to put forward the \devops{} initiative.

\interviewee{I3} highlighted the risks related to \devops{} adoption and suggested strategies to overcome them.
For him, a very close plan should be taken, presenting a history of broad \devops{} use in the organization.
Only sell the idea that \devops{} solve all the organization areas does not work.
On the other hand, the effective approach should present that the whole organization is already adopting \devops{}, 
there are already organization teams with experience in \devops{}, and there is evidence that many problems in projects were solved using \devops{}.
He also considered that \devops{} might be difficult and risky to adopt, requiring a high level of adaptation.

\interviewee{I4} reported the strategy employed by his organization in \devops{} adoption. 
He said \devops{} adoption resulted from a natural and spontaneous need of the development team.
There was a significant organization initiative for \devops{} adoption which concluded in the last year when a multidisciplinary working group was created to handle many problems that should be solved with \devops{} adoption.
There were sparse initiatives in some teams, but the main idea was to turn it into a corporate endeavor. 
The working group's goal was to make the \devops{} initiative officially in the company.
An effective plan was devised, which resulted in \devops{} adoption in a bottom-up way.

\section{Discussion}

This section presents an analysis of the interviewee's definitions related to the ones found in the SLR, highlighting differences and similarities. 

We generated a word cloud from the definitions gathered from the 28 documents selected in the SLR depicted in Figure~\ref{fig:concept-devops}.
The most relevant concepts resulting from the interviews were: agility, collaboration, continuous delivery, practices, process, improvements, and culture.
These concepts and their variations have high frequency when compared with the word cloud, which demonstrates, in a general view, that literature and industry converge. 
On the other hand, some items were mentioned a lot in the SLR but were not pointed out by the interviewees, such as software, quality, automation, teams, change, tools, principles, and approach.

The SLR brought 13 models or frameworks from which six are focused on \devops{} adoption, while the interviewees did not mention anyone with this focus.
It shows that the SLR finds are not commonly used on a large scale, or there is no recognition of its applicability in practice. 

The interviewees highlight the way for \devops{} is particular to each organization.
However, we argue that a model or framework should be used since they are defined from various previous experiences and industry reports.

Culture and automation principles are the most cited principles in SLR and by the interviewees.
However, several other principles are very relevant when considering the SLR but were not mentioned by the interviewees, such as Collaboration, Communication, and Agility.

Six practices were cited by the interviewees and found in the SLR: Continuous Deployment, Continuous Improvement, Continuous Testing and Automation, Automating Building, Infrastructure-as-Code, and Cloud Computing.

The majority of items reported by the interviewees were in the Difficulties perspective, which may be an obstacle for the advance of \devops{} adoption in organizations.

Table~\ref{tab:comparativo} presents a comparison between the number of items found in SLR and the Case Study for the eight identified perspectives.
Table~\ref{tab:comparativo2} presents a comparison of the reinforced items (\ie{}  appeared in SLR and Case Study) and new items (\ie{} emerged from the Case Study).

Since the SLR and Case Study are subjective processes, the presented numbers may have significant variance.
The results represent a snapshot in time, and they are not deterministic, \ie{} they are tight to the nature of an interpretative process.
So, we only compared and verified if there were convergence considering the findings of both methods application.
In this case, from the analysis presented in Table~\ref{tab:comparativo} and Table~\ref{tab:comparativo2}, 
we may deduct that there is a majority convergence among SLR and Case Study findings because there was empirical evidence of common elements.

\begin{table}[]
	\centering
	\caption{Comparative SLR and Case Study}
	\label{tab:comparativo}
	\begin{tabular}{|c|c|c|}
		\hline
		& \textbf{SLR} & \textbf{\begin{tabular}[c]{@{}c@{}}Case \\ Study\end{tabular}} \\ \hline
		Concepts    & 28           & 4                                                                  \\ \hline
		Models      & 13           & 1                                                                  \\ \hline
		Principles   & 15           & 2                                                                  \\ \hline
		Practices     & 73           & 11                                                                 \\ \hline
		Difficulties & 40           & 31                                                                 \\ \hline
		Challenges     & 52           & 13                                                                 \\ \hline
		Benefits   & 70           & 21                                                                 \\ \hline
		Strategies  & 99           & 21                                                                 \\ \hline
		Total        & 390          & 104                                                                \\ \hline
	\end{tabular}
\end{table}

\begin{table}[]
	\centering
	\caption{Case Study Items}
	\label{tab:comparativo2}
	\begin{tabular}{|c|c|c|}
		\hline
		& \textbf{\begin{tabular}[c]{@{}c@{}}Reinforced \\ Items\end{tabular}} & \textbf{\begin{tabular}[c]{@{}c@{}}New \\ Items\end{tabular}} \\ \hline
		Concepts    & 0  & 4  \\ \hline
		Models      & 0  & 1  \\ \hline
		Principles   & 2  & 0  \\ \hline
		Practices     & 8  & 3  \\ \hline
		Difficulties & 24 & 7  \\ \hline
		Challenges     & 11 & 2  \\ \hline
		Benefits   & 16 & 5  \\ \hline
		Strategies  & 14 & 7  \\ \hline
		Total        & 75 & 29 \\ \hline
	\end{tabular}
\end{table}

\section{Conclusions}

This research aimed to elucidate the phenomenon of \devops{} adoption in Brazilian organizations that develop and holistically maintain the software. For this, a multimethod approach was carried out with two methodological research techniques. Firstly, a Systematic Literature Review to understand the phenomenon, find the gaps and thus bring the main perspectives in their nuances associated with the theme, thus bringing the academy's view. In a second step, a Case Study was carried out with 2 Brazilian companies, one large and one small, to observe and empirically verify how the phenomenon of \devops{} adoption presented itself, thus allowing comparison and complementation to what was found in the literature.

With the development of this research, it was possible to understand and visualize the \devops{} phenomenon from the following perspectives found in the literature and sequentially applied in the industry: its most relevant principles, the varied concepts and interpretations involved, the associated practices, the problems faced, the current challenges, the benefits identified and the strategies used. This research result brought a wide range of characteristics and views, both in the literature and in the industry, on \devops{} manifestation in organizations. 

The SLR brought 390 items, artifacts, or characteristics related to the adoption of \devops{} were identified. The case study reinforced or validated 75 items and revealed 29 new items, expanding the literature findings.

Literature reports that the \devops{} initiative must be something \textit{top-down}, with support and sponsorship from top management to be successful in its adoption~\cite{bucena2017simplifying}~\cite{colomo2018case}~\cite{elberzhager2017agile}. However, in the case study, it was identified that the \textit{bottom-up} initiative proved to be fundamental for the successful \devops{} adoption in one of the companies. While confronting what was recommended in SLR, this fact converges with the idea that \devops{} came from the practical community and has proven itself over time as something beneficial and transformative for software engineering.

In short, it is possible to conclude that \devops{} manifests itself in many different ways and perspectives in organizations that develop software because its adoption is something transformative for the entire company, involving people, processes, and tools. Besides, it brings a robust cultural change, the need for new skills, greater collaboration, and organizational restructuring. It is not trivial, as \devops{} adoption requires a series of requirements and a very well-designed and structured plan to occur without trauma or at a high cost. This is mainly due to the involved human factor because without people being convinced, the journey can go wrong at all hierarchical levels.

Because it is a relatively new topic and comes from practice, associations need to mature the idea well, discussing it internally before deciding on \devops{} adoption. There is no magic guide or formula ready for doing it, a very particular journey for each company. It is pretty challenging to find the ideal path that can connect people, adjust processes and adopt tools combined and not necessarily in that order. In other words, it is a new way of working, requiring a lot of collaboration between teams, support from top management, motivation from both the company and employees, clarity about the objectives, a well-structured plan, with phases, and participatory adoption of ideal tools in the assembly of the \textit{pipeline}, well-structured documents and processes and continuous review for improvements.

Below is a list of possible future works to be considered from this research:

\begin{itemize}
	\item Use the Multivocal Literature Review methodology, according to the study of ~\cite{garousi2019guidelines}, to systematically aggregate publications related to \textit{Grey Literature}, thus taking a stronger view of the industry and practical community;
	\item Seek generalization of results using a quantitative research approach, such as a \textit{Survey} to validate the perspectives raised in other development environments;
	\item Conduct new case studies, in organizations from different branches of industry, in different countries, to seek and aggregate more cultural and organizational contexts;
	\item Apply new research focused on how the manifestations of the principles influence \devops{} adoption, with emphasis on culture and collaboration. There is relevant research on the topic developed that can be used as a reference, as \cite{humble2018accelerate};
	\item Adapt or build a model or \textit{framework} of best practices for adopting \devops{}, in addition to testing it in organizations, based on the results of this research.
\end{itemize}

\bibliographystyle{unsrtnat}
\bibliography{devops}  






\end{document}